\documentstyle[aps,psfig,multicol]{revtex}
\newcommand{\ignore}[1]{}
\begin{document}
\def\be{\begin{equation}}
\def\ee{\end{equation}}
\def\ba{\begin{eqnarray}}
\def\ea{\end{eqnarray}}

\title{Semiclassical Accuracy in Phase Space for Regular and Chaotic Dynamics}
\author{L. Kaplan}
\address{Department of Physics, Tulane University, New Orleans, LA 70118 USA}
\maketitle
\begin{abstract}
A phase-space semiclassical approximation valid to $O(\hbar)$ at short times is
used to compare semiclassical accuracy for long-time and stationary observables
in chaotic, stable, and mixed systems. Given the same level of semiclassical
accuracy for the short time behavior, the squared semiclassical error in the
chaotic system grows linearly in time, in contrast with quadratic growth in the
classically stable system. In the chaotic system, the relative squared error at
the Heisenberg time scales linearly with $\hbar_{\rm eff}$, allowing for
unambiguous semiclassical determination of the eigenvalues and wave functions
in the high-energy limit, while in the stable case the eigenvalue error always
remains of the order of a mean level spacing. For a mixed classical phase
space, eigenvalues associated with the chaotic sea can be semiclassically
computed with greater accuracy than the ones associated with stable islands.
\end{abstract}

\begin{multicols}{2}

\section{Introduction}

Semiclassical methods have a long history traceable to the very beginnings of
the ``old quantum theory" and serve two interrelated purposes in many areas of
physics. First, semiclassical methods provide valuable approximation techniques
in situations where a full quantum calculation is either impossible or
unnecessary. Equally importantly, semiclassical methods provide a link between
quantum results and our classical intuition, and allow us to separate physical
behavior that is due simply to classical paths and their interference from
behavior that is attributable to non-classical processes such as tunneling or
diffraction.

For strongly chaotic systems, purely {\it classical} calculations in $d$
dimensions that ignore phase effects must break down at the mixing time or log
time $T_{\rm log} \sim \lambda^{-1} \ln N \sim \lambda^{-1} (d-1)\ln \hbar_{\rm
eff}^{-1}$, where $\lambda$ is the maximal Lyapunov exponent of the classical
dynamics and $N \sim \hbar_{\rm eff}^{-(d-1)}$ is the effective dimension of
the accessible Hilbert space, or the size of the accessible classical phase
space in Planck cell units. This breakdown of {\it classical}--quantum
correspondence occurs because beyond the mixing time, multiple classical paths
connect a generic initial state to a generic final state, and interference
effects become $O(1)$. On the other hand, in a series of papers, Heller and
coworkers showed that {\it semiclassical} calculations in chaotic systems,
which include the effect of interference between distinct classical paths, can
follow the quantum propagator at times well beyond the mixing
time~\cite{earlyheller}. An estimate for the breakdown time scale of
semiclassical--quantum correspondence was obtained by quantifying the effects
of caustics for a stadium billiard~\cite{hellerstadium}.

Over the past decade, significant light has been shed on the issue of
semiclassical accuracy and its breakdown in diverse chaotic and regular
systems. For example, Boasman has used a semiclassical approximation to the
boundary integral method to obtain a semiclassical spectrum for two-dimensional
chaotic billiards, observing an overall semiclassical spectral shift as
compared with the exact quantum spectrum, in addition to small random
fluctuations~\cite{boasman}. On the other hand, Prosen and Robnik have shown
the complete failure of torus quantization to reproduce the spectra of
two-dimensional {\it integrable} billiards, such as the circle
billiard~\cite{prorob}, suggesting that integrability may in some cases lead to
an increase of semiclassical errors; Rahav {\it et al.} have obtained more
recent results consistent with this conclusion~\cite{rahav}. Primack and
Smilansky were among the first to analyze semiclassical accuracy for
three-dimensional chaotic systems, focusing on including corrections to the
state-counting function beyond the leading Weyl term~\cite{prsm}.

Main and collaborators have developed the powerful harmonic inversion technique
for accurate and efficient semiclassical calculations of energies, resonances,
and matrix elements~\cite{main1}. This technique, as well as the earlier
cycle-expansion method~\cite{cycle} were applied recently to the four-sphere
scattering problem, demonstrating a high degree of accuracy at a greatly
reduced computational cost compared with brute-force quantum
calculations~\cite{main2}. Another promising recent approach, put forward by
Vergini and coworkers, involves the accurate construction of quantum
eigenstates as linear superpositions of ``scar functions" associated with short
unstable periodic orbits~\cite{vergini}.

In the time domain, statistical arguments concerning the propagation of
semiclassical errors have shown that semiclassical error in chaotic systems
accumulates incoherently, and thus the squared error typically grows only
linearly with time, in contrast with quadratic growth for the regular
case~\cite{ltsc}. Transforming to the the energy domain, this implies that
semiclassical methods are generically more accurate for computing wave
functions and eigenvalues for chaotic systems than for regular ones, in the
$\hbar_{\rm eff} \to 0$ (or high energy) limit. In particular, for $d=2$,
analytical arguments and numerical tests show that eigenvalues can be
semiclassically resolved with great accuracy for chaotic systems, for
sufficiently small $\hbar_{\rm eff}$, while in the regular case even the order
of eigenvalues cannot be unambiguously determined semiclassically. This result
has been related to the reduction of the quantization ambiguity in chaotic
systems~\cite{quamb} and to the slower decay of fidelity in the presence of
strong chaos (as long as the perturbation has non-zero diagonal matrix
elements in the basis of the unperturbed system)~\cite{fidelity}.

Both the theoretical analysis and the numerical tests in Ref.~\cite{ltsc} were
performed for semiclassical evolution in the position representation, i.e., for
the Van-Vleck--Gutzwiller propagator~\cite{gutz}. Although well-suited for the
model systems treated in that work, position-representation semiclassics
suffers in general from the problem of proliferation of caustics, which
eventually dominate the semiclassical propagator~\cite{hellerstadium}. The
problem becomes particularly acute when one attempts to compare semiclassical
dynamics in hard chaotic systems with that in a regular system or in a mixed
phase space. Semiclassical calculations in a phase space basis are more natural
from the point of view of classical--quantum correspondence and have the
inherent advantage of allowing direct comparison between time evolution in
chaotic, regular and mixed systems, without the result being overwhelmed by the
problem of position-space or momentum-space caustics.

The aim of this paper is to improve our understanding of semiclassical accuracy
in a phase space representation, as a function of time and $\hbar_{\rm eff}$,
and to directly compare the behavior of the semiclassical error in chaotic,
regular, and mixed systems. The organization is as follows. In
Section~\ref{method} we briefly present the model and the method used for
performing semiclassical and quantum calculations in phase space. Theoretical
expressions for semiclassical accuracy in chaotic systems are presented in
Section~\ref{chaodyn}, along with supporting numerical data from the model
system. In addition to generalizing the analysis of Ref.~\cite{ltsc} from
position space to a phase space represenation, we explictly test the prediction
of Ref.~\cite{ltsc} concerning the linear growth with time of the mean squared
semiclassical error, as well as the prediction of linear decrease with $\hbar$
of the error at the Heisenberg time. This is followed by a similar analysis for
regular and mixed systems, in Sections~\ref{regdyn} and \ref{mixeddyn},
respectively. We find that the behavior of the semiclassical error at the
energy scale of a mean level spacing corresponds well to the effect of the
quantization ambiguity on the spectrum, which was previously studied in
Ref.~\cite{quamb}. Finally, Section~\ref{summary} summarizes the results and
presents an outlook for the future.

\section{Model and Method}
\label{method}

Although our theoretical analysis applies quite generally to two-dimensional
noninteracting systems, we simplify the numerical simulations by focusing on
one-dimensional kicked maps, which share most scaling and other physical
properties of this class of systems~\cite{maps}. The discrete-time map can be
regarded as a Poincar\'e surface of section of a two-dimensional system with
Hamiltonian dynamics. Specifically, we parametrize the one-step map on a
toroidal classical phase space $(q,p) \in [0,1) \times [0,1)$ as
\begin{eqnarray}
p_0 \rightarrow p_1&=&p_0 - V'(q_0) \;\;{\rm mod} \;\; 1\nonumber \\
q_0 \rightarrow q_1& = & q_0+ T'(p_1)\;\; {\rm mod} \;\;1 \,,
\end{eqnarray}
where
\begin{eqnarray}
V(q)&=& -{1 \over 2} m_q q^2 - {K_q \over (2 \pi)^2} \sin(2\pi q) \nonumber \\
T(p)&=& {1 \over 2} m_p p^2 + {K_p \over (2 \pi)^2} \sin(2\pi p) \,.
\label{vandt}
\end{eqnarray}
The dynamics is iterated to obtain classical evolution over many kicks (or many
bounces in the corresponding two-dimensional Hamiltonian system). For values
$m_q=m_p=1$ and $0< |K_q|, |K_p|<1$, for example, we obtain the purely chaotic
perturbed cat map, or kicked inverted oscillator, while for $m_q=-1$, $m_p=1$
and small $K_q$, $K_p$, the dynamics is predominantly regular, corresponding to
a kicked regular oscillator. Parameters $K_q$ and $K_p$ are essential to
introduce nonlinearity into the dynamics (if $K_q=K_p=0$, the semiclassical
propagator is exact, for any $m_q$ and $m_p$). The above four parameters can
also be adjusted to vary the Lyapunov exponent in the chaotic regime, or to
study a mixed phase space, as we will see below in Section~\ref{semacc}.

The one-step quantum evolution matrix for the above system takes the very
simple form \begin{equation}
\hat U_1= \exp{[-i\hat T(\hat p)/\hbar]} \exp{[-i \hat V(\hat q)/\hbar]} \,,
\end{equation}
which again may be iterated or diagonalized to obtain long-time or stationary
behavior, $\hat U_t = [\hat U_1]^t$. As discussed in the introduction, we will
apply this propagator to Gaussian wave packets (or coherent states) centered at
phase space points ($q_k$, $p_k$):
\begin{equation}
\label{gaussian}
\phi_k(q) = {\cal N} \exp {\left[- (q-q_k)^2/2\hbar +i
p_k(q-q_k)/\hbar\right]}\\,
\end{equation}
where ${\cal N}$ is a normalization constant.

Unfortunately, there is not a unique and universally used semiclassical
approximation for wave packet evolution, analogous to the Van-Vleck--Gutzwiller
expression in position or momentum space. Several methods have been proposed
that differ in both the order (in $\hbar_{\rm eff}$) of the semiclassical error
at fixed time $t$ and in the numerical size of that error. The so-called
``thawed Gaussian" approximation, for example, allows the shape of the Gaussian
wave packet to change as it evolves under a locally quadratic potential
$V(q)$~\cite{thawed}. An alternative approach uses ``frozen" or unspreading
wave packets~\cite{hk}. Another coherent state method retains the stationary
phase idea of the Van-Vleck--Gutzwiller propagator but extends dynamics into
complex phase space~\cite{hhl}. It is possible instead to work in complex time
while retaining real initial conditions in phase space~\cite{miller}.

In the present work, we are not interested in reducing the numerical size of
the semiclassical error but only in understanding its scaling properties with
$t$ and $\hbar_{\rm eff}$, for regular and chaotic systems. For this reason, we
will choose what is a convenient method for our purposes, noting that the
results would hold for any semiclassical approximation valid to the same order
in $\hbar_{\rm eff}$. We essentially use a variation of the thawed Gaussian
method, extended to next-to-leading order in $\sqrt{\hbar_{\rm eff}}$, and then
calculate semiclassically the overlaps of the time-evolved ``thawed" Gaussians
with the Gaussians in the original basis~\cite{baker}.

Specifically, we start with a (non-orthogonal) complete set of $N=1/2\pi\hbar$
Gaussians $\phi_j$ of the form given in Eq.~(\ref{gaussian}), with the center
points $(q_j,p_j)$ offset slightly from a rectangular grid to reduce numerical
instabilities. The semiclassical overlap matrix
\begin{equation}
A_0(j,k) = \langle \phi_j | \phi_k \rangle _{\rm SC}
\end{equation}
is obtained analytically by Gaussian integration. To evaluate the $t-$step
semiclassical propagator $A_t(j,k)$ between initial Gaussian $\phi_k$ and final
Gaussian $\phi_j$, we find real classical trajectories from $(q_0,p_0)$ to
$(q_t,p_t)$ in time $t$ that minimize
$(q_0-q_k)^2+(p_0-p_k)^2+(q_t-q_j)^2+(p_t-p_j)^2$, i.e., all trajectories that
start near the center of Gaussian $k$ and end near the center of Gaussian $j$
after $t$ steps. Of course for fixed $t$ and sufficiently small $\hbar$ ($t <
T_{\rm log} \sim \lambda^{-1} \log \hbar^{-1}$), there will be at most one such
trajectory, and in principle that is all we need even for our long-time
analysis, as will be seen below. In practice however, for finite values of
$\hbar$ we include all contributing trajectories. For each trajectory, the
potential $V(q)$ is expanded to {\it third} order around the starting position
of the trajectory, $q_0$. When this potential is applied to the original
Gaussian $\phi_k$, we obtain a wave packet of the form
\begin{eqnarray}
\label{newpacket}
& & \exp{\left[ a+ b(q-q_0)+c(q-q_0)^2+d(q-q_0)^3 \right]} \\ \nonumber
& = &
\exp{\left [a+ b(q-q_0)+c(q-q_0)^2 \right]} \\ \nonumber & \times &
\left[1+d(q-q_0)^3
+O(\hbar)\right] \,,
\end{eqnarray}
where $a$, $b$, $c$, and $d$ are complex numbers of order $\hbar^{-1}$, and
therefore $q-q_0$ is $O(\hbar^{1/2})$. We note that an ``extended"
semiclassical dynamics~\cite{ps}, which truncates the expansion of the
Hamiltonian at third order rather than second order is needed to keep the error
in the one-step phase space propagator at $O(\hbar)$, consistent with the error
in the Van-Vleck--Gutzwiller propagator in position space~\cite{gutz}.

The wave packet of Eq.~(\ref{newpacket}) may now be rewritten, via Fourier
transform, as a momentum space wave packet having the same form but expanded in
powers of $p-p_1$ instead of $q-q_0$. The kinetic term $T(p)$ of the
Hamiltonian may now be applied, again expanded to third order in $p-p_1$. Then,
the packet is Fourier transformed back to position space and the procedure is
repeated $t$ times. At the end of $t$ steps, we may analytically find the
overlap between the semiclassically evolved $t-$step wave packet $\phi_{k,{\rm
SC}}(t)$, still having the form of Eq.~(\ref{newpacket}), and the final
Gaussian wave packet $\phi_j$ to obtain the semiclassical propagator
$A_t(j,k)$. If several classical paths lead from the vicinity of $\phi_k$ to
the vicinity of $\phi_j$ in time $t$, their contributions must be summed to
produce the semiclassical amplitude $A_t(j,k)$, just as in the Gutzwiller
expression. As we will see in Section~\ref{chaodyn}, for a chaotic system the
long-time semiclassical propagator may be arbitrarily well approximated (in the
$\hbar \to 0$ limit) using only the matrix $A_t$ for $ 1 \ll t \ll T_{\rm
log}$, where at most one path contributes to each matrix element. However, as
we are dealing with finite $\hbar$ in our numerical simulations, we will always
use the sum over all classical paths in numerical calculations.

\section{Semiclassical Accuracy}
\label{semacc}

\subsection{Chaotic Dynamics}
\label{chaodyn}

As discussed previously in the context of position-space semiclassical
propagation, direct comparison between quantum and semiclassical evolution at
long times for a chaotic system, or between quantum and semiclassical
stationary properties for such a system, faces the obstacle of the exponential
proliferation of classical paths~\cite{earlyheller}; an analogous problem of
exponential growth in the number of periodic orbits exists in the energy
domain~\cite{gutz,cycle,cycle2}. This proliferation seemingly makes long-time
semiclassical propagation in a classically chaotic system an exponentially
harder problem than the full quantum evolution, puts into question the
convergence of long-time semiclassical dynamics to any stationary behavior, and
prevents the comparison of semiclassical and quantum stationary properties for
small $\hbar$. The threefold difficulty can be addressed using the idea that
the Heisenberg uncertainty principle washes out information on scales below
$\hbar$, and thus the total amount of semiclassical information is finite for
all times and scales only as a power of $\hbar$. We can therefore collect,
consolidate, and iterate semiclassical amplitude on sub-$\hbar$ scales,
obtaining the full semiclassical long-time dynamics to arbitrary accuracy in
polynomial computation time. This ``semiclassical path consolidation" idea has
previously been used successfully to investigate long-time semiclassical
accuracy in the position representation for chaotic dynamics~\cite{baker,ltsc}
and to demonstrate the semiclassical nature of dynamical localization in one
dimension~\cite{scl}. Conceptually, the approach is similar to cycle expansion
methods in periodic orbit theory~\cite{cycle,cycle2}; however, no information
about periodic orbits is needed here. Instead of accounting for long-time
semiclassical behavior in terms of periodic paths up to period $\tau_{\rm
periodic} \sim T_{\rm log}$, we use {\it all} short paths up to length $\tau
\sim 1$.  In the following, we adapt the methods of Ref.~\cite{ltsc} to a phase
space representation, and refer the reader to that earlier paper for a detailed
discussion.

We begin by noting that although semiclassical dynamics is not multiplicative,
due to the fact that a concatenation of two stationary paths is in general not
stationary, we may nevertheless write
\begin{eqnarray}
A_{t_1+t_2}(j,k)& =& \sum_{\ell,\ell'} A_{t_2}(j,\ell')A_0^{-1}(\ell',\ell)
A_{t_1}(\ell,k) +O(\hbar) \nonumber \\
& = & \left[A_{t_2} A_0^{-1}A_{t_1} \right] (j,k)
+O(\hbar)\,,
\end{eqnarray}
where the $O(\hbar)$ error is due to the intermediate sums being done exactly
rather than by stationary phase, and the inverse of the semiclassical overlap
matrix $A_0$ is necessary due to non-orthogonality. In general, we may
approximate the true time-$t$ semiclassical propagator $A_t$ by evaluating the
exact semiclassical dynamics to some ``quantization time" $\tau$ and then
iterating the resulting matrix:
\begin{equation} A_{t,\tau} = A_{t \; {\rm mod} \; \tau} \left [A_0^{-1}
A_{\tau} \right ] ^{[t/\tau]} \,
\end{equation} where $[t/\tau]$ is the integer part of $t/\tau$. We may call
$A_{t,\tau}$ the ``$\tau$-semiclassical" propagator. For $\tau=1$,
$A_{t,\tau}$ is the one-bounce semiclassical quantization pioneered by
Bogomolny~\cite{bogoquant}. For a continuous-time system, the $\tau \ll 1$
limit is equivalent to quantum propagation via the Feynman path integral
approach. The exact time-$t$ semiclassical propagator, on the other hand, is
recovered in the opposite limit when the quantization time approaches $t$:
\begin{equation}
\label{atqconv}
A_{t,\tau} \to A_t \;\;\; {\rm as} \;\;\; \tau \to t \,.
\end{equation}
In Ref.~\cite{ltsc}, it was shown analytically and numerically that the error
$|A_{t,\tau}(j,k)-A_t(j,k)|^2$ falls off as $T_{\rm cl}/\tau$ in a chaotic
system, where $T_{\rm cl}$ is the time scale of classical correlations. This
implies that for $\tau \gg T_{\rm cl}$ the approximate semiclassical correlator
$A_{t,\tau}$ is closer to the exact semiclassical correlator than either is to
the quantum dynamics:
\[
\left | A_{t,\tau} - A_t \right| \ll \left |A_t - U_t \right| \,.
\]
Thus
\begin{equation}
\label{errconv}
\left | A_{t,\tau} - U_t \right | \stackrel{\tau \gg T_{\rm cl}}{\longrightarrow}
\left |A_t - U_t \right| \,
\end{equation}
allowing for an unambiguous determination of the error in the true
semiclassical dynamics $A_t$ at time $t$ using $A_{t,\tau}$ and permitting a study of
the breakdown of the semiclassical approximation at long times $t$ where
performing an exact sum over $O \left (e^{\lambda t}\right)$ classical paths is
impractical or impossible.

\begin{figure}
\centerline{
\psfig{file=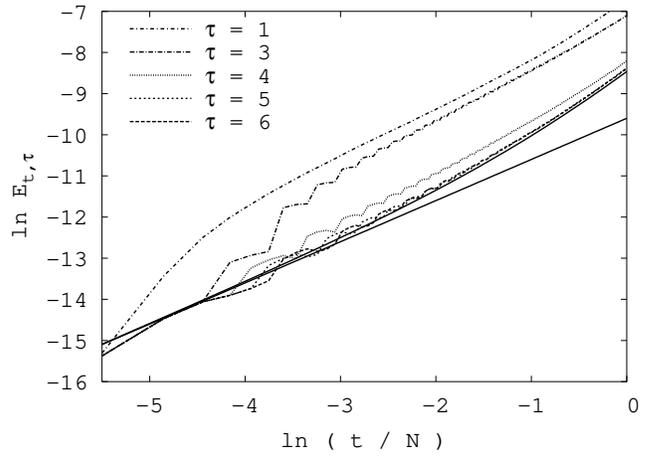,angle=270,width=3.5in}}
\vskip 0.2in
\caption{The mean squared $\tau$-semiclassical error $E_{t,\tau}$ in a chaotic
system is plotted as a function of $t/N$ for semiclassical parameter $N=256$
and for several values of the quantization time $\tau$. For $\tau \gg 1$,
$E_{t,\tau}$ is a reliable proxy for the true semiclassical error $E_t$. The
classical system parameters for Eq.~(\ref{vandt}) are $m_q=m_p=1$;
$K_q=K_p=1/2$. The lower and upper solid curves represent the theoretical
predictions for $E_t$ given by Eqs.~(\ref{et1}) and (\ref{et2}), respectively,
with $C_1=0.017$ and $C_2=0.037$.}
\label{fig_tim}
\end{figure}

To confirm the convergence of the iterated propagator $A_{t,\tau}$ to the true
long-time semiclassical propagator $A_t$ for semiclassical dynamics in phase
space, and specifically the convergence of the semiclassical error in
accordance with Eq.~(\ref{errconv}), we first compute, as a function of time
$t$, the average $\tau-$semiclassical error defined as:
\begin{eqnarray}
E_{t,\tau} &=& ||A_{t,\tau}-U_t||^2= Tr
[A_{t,\tau}-U_t]^\dagger[A_{t,\tau}-U_t] \nonumber \\ &=&
\sum_{j,k}\left|A_{t,\tau}(j,k)-U_t(j,k)\right|^2 \,.
\end{eqnarray}
The results are shown in Fig.~\ref{fig_tim} for a chaotic kicked map defined by
parameters $m_q=m_p=1$ and $K_q=K_p=1/2$, with semiclassical parameter $N=256$.
We notice the relatively poor agreement between the iterated semiclassical
calculation for $\tau=1$ and similar calculations for larger quantization times
$\tau$. We also note that using the iterated propagator with short quantization
time $\tau$ overestimates the true size of the semiclassical error. At the same
time, we observe rapid convergence of $E_{t,\tau}$ as $\tau \gg 1$, with the
$\tau=5$ and $\tau=6$ curves lying almost on top of one another. Thus, the
$\tau$-semiclassical error $E_{t,\tau}$ appears to be rapidly approaching the
true semiclassical error
\begin{equation}
E_{t} =\sum_{j,k}\left|A_t(j,k)-U_t(j,k)\right|^2 \,.
\end{equation}

We are now ready to investigate the semiclassical error $E_t$ as a function of
time $t$ and semiclassical parameter $N$. For a chaotic system, we may assume
that the errors associated with the semiclassical approximation add
incoherently as long as the times at which the errors occur are separated by at
least the classical time scale $T_{\rm cl}$~\cite{ltsc}. Since the squared
error in the semiclassical approximation over one time step is
$E_1 = O(h^2)=O(1/N^2)$, we obtain
\begin{eqnarray}
\label{et1}
E_t &=& C_1 h^2 t \nonumber \\
&=& h C_1\left({t \over N}\right)\,,
\end{eqnarray}
where $C_1 \sim T_{\rm cl}$ is a system-dependent constant and we take $t=1$ to
correspond to one period of the kicked map. The linear growth of the error
predicted by Eq.~(\ref{et1}) breaks down at times comparable to the Heisenberg
time, where we must include an additional error term that is diagonal in the
eigenbasis of the true quantum propagator $U_1$~\cite{ltsc}. The error
associated with diagonal matrix elements adds coherently, leading to quadratic
growth of the cumulative error in time. However, the fraction of diagonal
matrix elements scales as $h = 1/N$. Eq.~(\ref{et1}) must therefore be modified
to read
\begin{eqnarray}
\label{et2}
E_t &=& C_1 h^2 t +C_2 h^3 t^2\nonumber \\
&=& h \left[C_1 \left({t \over N}\right)+ C_2\left({t
\over N}\right)^2 \right]\,.
\end{eqnarray}

The data in Fig.~\ref{fig_tim} for $\tau \ge 4$ shows good agreement with the
prediction of Eq.~(\ref{et2}), which is indicated by the upper solid curve. The
linear growth indicated by Eq.~(\ref{et1}), shown as the lower solid line,
is valid for times $t$ short compared with the Heisenberg time $N$.

\begin{figure}
\centerline{
\psfig{file=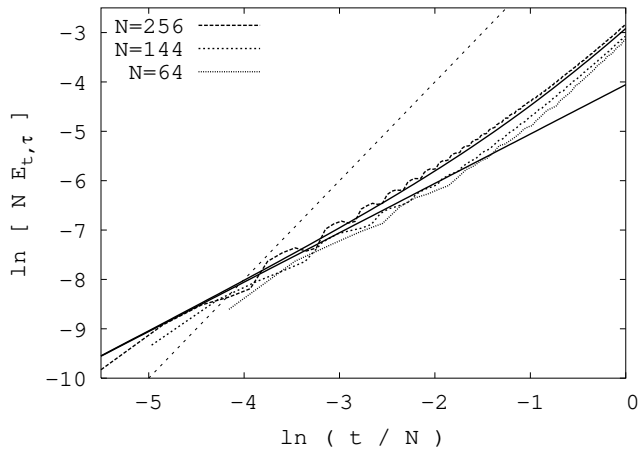,angle=270,width=3.5in}}
\vskip 0.2in
\caption{The mean squared $\tau$-semiclassical error $E_{t,\tau}$ for a chaotic
system is plotted as a function of $t/N$ for $\tau =5 \gg 1$ and several values
of the semiclassical parameter $N=1/h=64$, $128$, $256$. The classical system
parameters are the same as in the previous figure. The lower and upper solid
curves represent the theoretical predictions of Eqs.~(\ref{et1}) and
(\ref{et2}), respectively. The dotted line indicates the predicted growth of
the error for a system with {\it regular} dynamics, $E_t \sim t^2$ (see
Section~\ref{regdyn}), and is shown to emphasize the qualitatively different
behavior.}
\label{fig_timn}
\end{figure}

In Fig.~\ref{fig_timn}, we confirm the behavior predicted by Eqs.~(\ref{et1})
and (\ref{et2}) as we vary the semiclassical parameter $N=1/h$. In this figure,
the error $E_{\tau,t}$ has been scaled by a factor of $N$ to make the curves
at different values of $N$ approximately coincide and to emphasize that the
error at a fixed fraction of the Heisenberg time is falling off as $h \sim 1/N$
in the semiclassical limit $h \to 0$.

Specifically, we may ask about the size of the semiclassical error at the
Heisenberg time itself, i.e. at $t/N=1$, which corresponds to the right edge of
the graph in Figs. \ref{fig_tim} and \ref{fig_timn}. The scaling of the error
at the Heisenberg time determines the feasibility of semiclassically computing
individual eigenstates and eigenvalues in the limit of small $\hbar_{\rm eff}$,
corresponding physically to the high-energy limit $E \gg E_{\rm gs}$. Based on
Eq.~(\ref{et2}), we predict the error at the Heisenberg time to be proportional
to $h$:
\begin{equation}
E_{t=N} = h \left [ C_1 +C_2 \right] \,.
\label{eth}
\end{equation}

\begin{figure}
\centerline{
\psfig{file=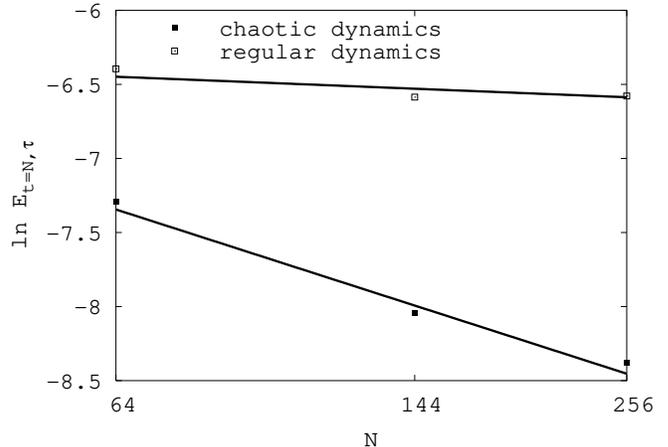,angle=270,width=3.5in}}
\vskip 0.2in
\caption{
The semiclassical error at the Heisenberg time, $E_{t=N,\tau}$ is plotted for
$\tau=5 \gg 1$, and for three values of the semiclassical parameter $N=1/h$.
Black squares correspond to the chaotic system of the previous two figures,
while white squares correspond to the regular system of Section~\ref{regdyn}.
The straight lines are fits to the power-law form $E_{t=N} = \alpha
N^{-\beta}$, with the best fit giving $\beta_{\rm chaotic}=0.8$ and $\beta_{\rm
regular}=0.1$, to be compared with the theoretical predictions $\beta_{\rm
chaotic}=1$ (Eq.~(\ref{eth})) and $\beta_{\rm regular}=0$
(Eq.~(\ref{ethreg})).}
\label{fig_eth}
\end{figure}

This prediction is tested in Figure~\ref{fig_eth}, where the black squares
represent the numerical data and the corresponding solid line is a best fit to
a power-law form, $E_{t=N}=a h^\beta= a N^{-\beta}$, with $\beta \approx 0.8$.
This is to be compared with the asymptotic prediction $\beta=1$ for $h \to 0$.
The falloff in the error with $N$ shows that individual eigenstates and
eigenvalues may be determined with ever improving accuracy as $N \to \infty$.
As we will find in the following section, this is in contrast with the
situation for systems with regular classical dynamics (see also the white
squares in Fig.~\ref{fig_eth}).

The semiclassical spectrum and semiclassical eigenstates can be obtained in
principle by computing the semiclassical propagator $A_t$ for long times and
transforming into the energy domain. However, since the semiclassical
propagator $A_t$ at long times becomes approximately
multiplicative~\cite{ltsc},
\begin{equation}
A(t+1) \approx A_\ast A(t)
\end{equation}
for some constant matrix $A_\ast$, it is much more convenient to diagonalize
$A_\ast$ directly to obtain the semiclassical eigenvalues and wave functions.
We emphasize that $A_\ast$ is neither the quantum evolution matrix $U_1$ nor
the semiclassical evolution matrix $A_1$ for one time step, but is instead the
effective one-step semiclassical propagator that describes semiclassical
evolution at long times, and thus the stationary behavior of the semiclassical
dynamics~\cite{ltsc}. In practice, we may obtain $A_\star$ as the limit
\begin{eqnarray}
A_\ast &=& \lim_{\tau \to \infty} A_{\ast,\tau} \nonumber \\
&=& \lim_{\tau \to \infty} A(\tau+1)\left[ A(\tau)\right]^{-1} \,.
\label{astarlimit}
\end{eqnarray}

\begin{figure}
\centerline{
\psfig{file=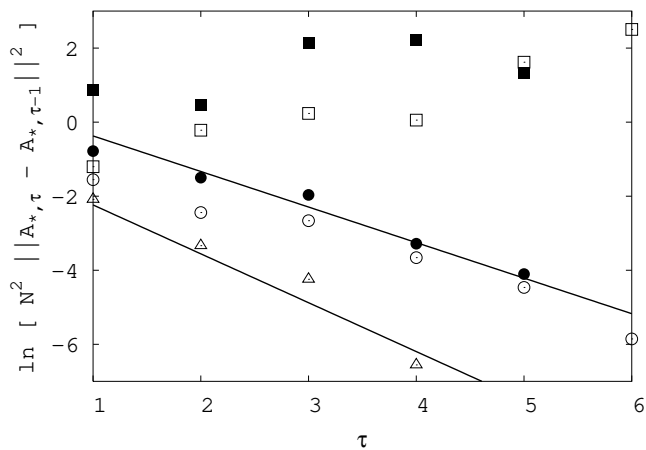,angle=270,width=3.5in}}
\vskip 0.2in
\caption{
The convergence of the finite-time approximation to the long-time one-step
semiclassical propagator $A_\ast$ is studied for several systems and different
values of the semiclassical parameter $N=1/h$ (see Eq.~(\ref{astarlimit})). The
circles represent data for the same system parameters that were used in the
previous three figures, white and black circles corresponding to $N=64$ and
$N=144$, respectively. White triangles represent data for $N=64$ with an
alternative set of parameters: $m_q=2$, $m_p=1$, $K_q=K_p=1/2$ in
Eq.~(\ref{vandt}), having a larger Lyapunov exponent. The solid lines for the
two systems are the predictions of Eq.~(\ref{astarerror}), with
$\lambda=\cosh^{-1}{(3/2)}=0.96$ and $\lambda=\cosh^{-1}{(2)}=1.32$,
respectively. The white and black squares ($N=64$ and $N=256$, respectively)
represent data for the regular dynamics discussed in Section~\ref{regdyn},
where no convergence with $\tau$ is predicted or observed.}
\label{fig_aconv}
\end{figure}

As discussed in Ref.~\cite{ltsc}, the convergence $A_{\ast,\tau} \to
A_\ast$ is exponentially fast in $\tau$, at least for the position space
semiclassical propagator:
\begin{equation}
||A_{\ast,\tau}-A_\ast||^2 \sim h^2 e^{-\lambda \tau} \,.
\label{astarerror}
\end{equation}
In Fig.~\ref{fig_aconv}, we verify this convergence in the case of the phase
space semiclassical propagator for two different values of $N=1/h$ (white and
black circles). The rate of convergence $\lambda$ is consistent with
the classical value of the Lyapunov exponent, and is independent of $\hbar$.
The white triangles correspond to an example with a larger Lyapunov exponent
($m_q=2$, $m_p=1$, $K_q=K_p=1/2$ in Eq.~(\ref{vandt})), where the convergence
with $\tau$ is correspondingly faster. 

Exponentially fast convergence to $A_\ast$ with $\tau$ implies that the
semiclassical spectrum and semiclassical wave functions can be obtained with
very high accuracy using semiclassical dynamics for $t >1$ but still short
compared to the Heisenberg time $t=N$ or even the log time $T_{\rm mix}$. In
other words, all the information needed to calculate long-time or stationary
semiclassical properties is already contained in the short-time classical
behavior, well before before interference effects become relevant.

The stationary semiclassical spectrum and wave functions can now be compared
with their quantum analogues. From the linear scaling with $h$ of the error in
the time evolution at the Heisenberg time, Eq.~(\ref{eth}), which has been
tested above in Fig.~\ref{fig_eth}, we can deduce that the mean squared error
in the eigenvalues must also scale linearly with $h$, ignoring a possible
overall shift in the spectrum~\cite{quamb} which is absent in the present
system due to symmetry. Thus,
\begin{equation}
\label{eigerror}
F={1 \over N} \sum_{i=1}^N {{\left(\epsilon_{i,{\rm SC}}-\epsilon_i\right)^2}
\over \Delta^2} \sim h = {1 \over N} \,,
\end{equation}
where the $\epsilon_i$ and $\epsilon_{i,{\rm SC}}$ are the quantum and
semiclassical eigenvalues, and $\Delta$ is the mean level spacing. In practice,
this improvement in the semiclassical approximation for individual eigenvalues
as $h \to 0$ is difficult to measure due to numerical errors. For example, for
the same chaotic system discussed previously ($m_q=m_p=1$, $K_q=K_p=1/2$), $F$
is already $1.3 \cdot 10^{-5}$ for $N=36$.

\subsection{Regular Dynamics}
\label{regdyn}

We may easily change parameters in Eq.~(\ref{vandt}) to obtain fully or almost
fully stable
classical dynamics and then repeat the semiclassical calculations and analysis
of Section~\ref{chaodyn}. We choose $m_p=1$, $m_q=-1$, $K_p=K_q=0.1$. The small
nonlinearity parameters $K_p$ and $K_q$ have been selected to reduce the
semiclassical error in the short-time propagator; as we will see below, the
semiclassical error grows much faster with time here than in the chaotic case.

\begin{figure}
\centerline{
\psfig{file=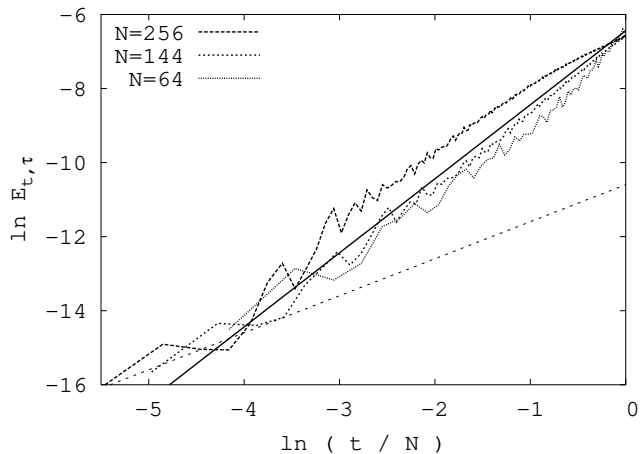,angle=270,width=3.5in}}
\vskip 0.2in
\caption{The mean squared $\tau$-semiclassical error $E_{t,\tau}$ for a system
with regular dynamics is plotted as a function of $t/N$ for $\tau =5 \gg 1$ and
several values of the semiclassical parameter $N=1/h=64$, $128$, $256$. The
classical system parameters are $m_p=1$; $m_q=-1$; $K_q=K_p=0.1$. The solid
curve represents the theoretical prediction of Eq.~(\ref{etreg}), with
$C=0.0016$. The dotted line corresponds to linear growth of the error with
time, $E_t \sim t$, applicable to the chaotic case only (see Eq.~(\ref{et1})
and Fig.~\ref{fig_timn}), and is shown to emphasize the very different scaling
behavior in the case of regular dynamics.}
\label{fig_timnreg}
\end{figure}

In a system with regular dynamics, a typical classical trajectory repeatedly
visits the same regions of phase space, and errors in the semiclassical
approximation are expected to add coherently~\cite{quamb}. Thus, in contrast
with the chaotic case, the squared difference between the time evolution matrix
for quantum dynamics and its semiclassical counterpart is expected to grow
quadratically with time:
\begin{eqnarray}
\label{etreg}
E_t &=& C h^2 t^2 \nonumber \\
&=& C\left({t \over N}\right)^2\,,
\end{eqnarray}
where $C$ is a classical constant that depends on the nonlinearity of the
system, as well as on the typical number of kicks needed for a typical
classical trajectory to return to the vicinity of its starting point. This
quadratic growth of the error, even at times short compared to the Heisenberg
time $N$, is to be contrasted with the result of Eq.~(\ref{et2}) for a fully
chaotic system. The prediction of Eq.~(\ref{etreg}) is tested in
Fig.~\ref{fig_timnreg}, where the quadratic growth is confirmed as well as the
predicted scaling with the semiclassical parameter $N=1/h$. Furthermore, the
growth of the semiclassical error with time is completely different in the
regular and chaotic case, as can be seen from the dotted lines in
Figs.~\ref{fig_timn} and \ref{fig_timnreg}.

For a regular system at the Heisenberg time $t=N$, we obtain an $\hbar_{\rm
eff}$-independent semiclassical error
\begin{equation}
E_{t=N} = C\,,
\label{ethreg}
\end{equation}
to be contrasted with the diminishing semiclassical error at the Heisenberg
time in the $\hbar_{\rm eff} \to 0$ limit for a chaotic system, as indicated by
Eq.~(\ref{eth}). The Heisenberg-time error for our regular system is plotted
for several values of $N$ in Fig.~\ref{fig_eth}. We note that the
Heisenberg-time semiclassical error is larger for the regular system as
compared with a chaotic system at the same value of $N$, despite the fact that
the one-step semiclassical error is larger in the chaotic case.

The $O(\hbar^0)$ error in the semiclassical evolution at the Heisenberg time,
as indicated by Eq.~(\ref{ethreg}), suggests that semiclassical eigenvalues and
wave functions, if they exist, do not approach the corresponding quantum
eigenvalues and wave functions in the $\hbar_{\rm eff} \to 0$ limit. Instead,
for a two-dimensional system with regular classical dynamics, the semiclassical
error in the eigenvalues is proportional to the mean level spacing, implying
that even the order of eigenvalues in the spectrum cannot be unambiguously
determined using semiclassical methods.

The problem, however, is more serious still, as the semiclassical dynamics for
a regular system does not in general approach a stationary behavior at long
times. We recall that for a chaotic system, the dynamics at long times
approaches multiplication by a constant matrix $A_\ast$, whose eigenvalues and
wave functions determine the stationary properties of the system. In contrast,
for a regular system, the convergence of Eq.~(\ref{astarlimit}) does not hold,
since the Lyapunov exponent vanishes. This lack of convergence is observed in
the squares plotted in Fig.~\ref{fig_aconv}, where it is seen that successive
approximations to $A_\ast$ differ from one another at $O(1/N^2)=O(\hbar_{\rm
eff}^2)$. In other words, the eigenvalues of the matrix defining semiclassical
evolution from time $t$ to $t+1$ and the eigenvalues of the matrix defining
semiclassical evolution from $t+1$ to $t+2$ differ from one another on the
scale of a mean level spacing, so no unique semiclassical spectrum can be
defined that describes the long-time semiclassical behavior.

We note that a system with regular dynamics may be separable, in which case one
may have a special set of coordinates for which semiclassical dynamics is exact
(just as semiclassics may be exact for special chaotic systems such as the cat
maps). The above results apply to the general situation where separability may
not hold, e.g., a pseudo-integrable system or a generic polygonal billiard, and
also to the separable case when the quantization is done in a set of
coordinates other than the ones for which the equations of motion separate.
Assuming the semiclassics is not exact, and independent of the initial size of
the semiclassical error, the semiclassical accuracy will progressively improve
in the $\hbar_{\rm eff} \to 0$ or high-energy limit as long as the Lyapunov
exponent $\lambda$ is nonzero, until eventually individual eigenvalues and wave
functions become semiclassically resolvable. In the case of zero Lyapunov
exponent, this improvement does not occur.

\subsection{Mixed Dynamics}
\label{mixeddyn}

Generic two-dimensional systems are neither fully regular nor fully chaotic,
and it is therefore of interest to study the issue of semiclassical-quantum
correspondence in the general regime of ``soft chaos." A mixed classical phase
space can be obtained using parameters $m_q=K_p=0$, $m_p=K_q=1$ in
Eq.~(\ref{vandt}); for this system approximately 48\% of phase space is
associated with the chaotic sea and the remainder consists of stable islands.
Based on our discussion in Sections~\ref{chaodyn} and \ref{regdyn} on the very
different behavior of semiclassical accuracy in chaotic and regular systems,
respectively, it is natural to ask whether semiclassical accuracy may vary with
initial conditions in the case of a mixed phase space.

\begin{figure}
\centerline{
\psfig{file=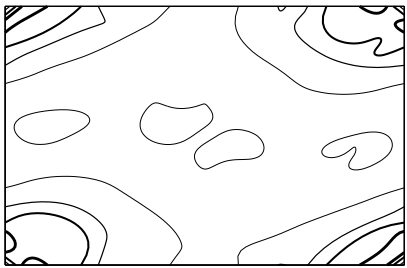,angle=270,width=4.0in}}
\centerline{
\psfig{file=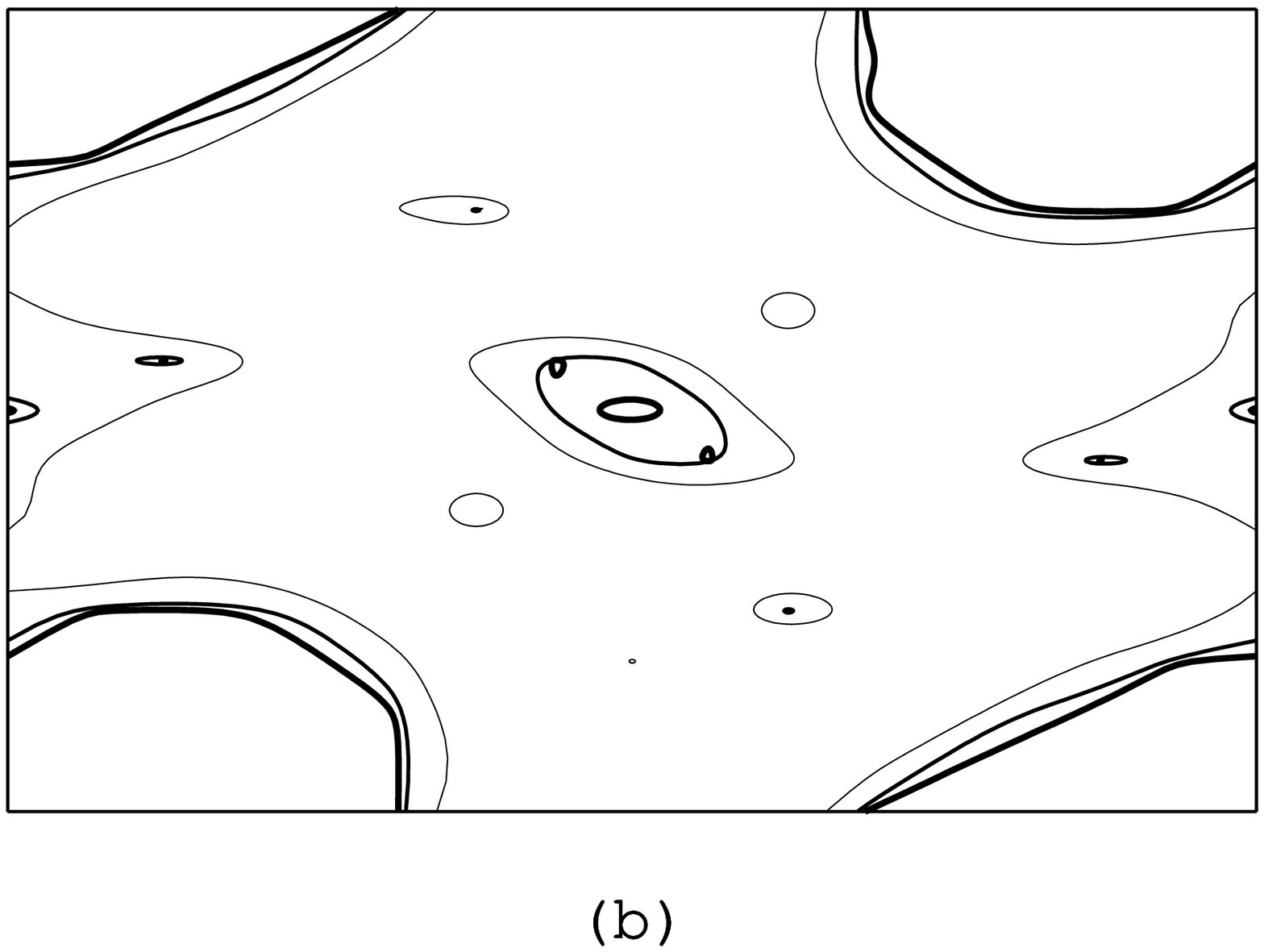,angle=270,width=4.0in}}
\caption{(a) The weighted semiclassical eigenvalue error $F_{\phi_k}$ is
plotted as a function of phase space location $(q_k,p_k)$ for semiclassical
parameter $N=256$. The contour curves correspond to $F=0.001$, $0.003$,
$0.005$, $0.007$ (the thickest curve indicates the largest error). The
semiclassical eigenvalues are obtained by diagonalizing $A_{\ast,0}$ in
Eq.~(\ref{astarlimit}). (b) For each wave packet $\phi_k$ used in (a), the
fraction of stable trajectories for that wave packet is calculated classically
and again plotted as a function of wave packet location. The contour curves
correspond to stable fractions of $0.6$, $0.9$, $0.975$ (the thickest curve
corresponding to the most stable region).
}
\label{fig_mixed}
\end{figure}

We define a local version of the mean squared eigenvalue error introduced in
Eq.~(\ref{eigerror}):
\begin{equation}
\label{loceigerror}
F_{\phi_k}=\sum_{i=1}^N |\langle \psi_i|\phi_k \rangle |^2
{\left(\epsilon_{i,{\rm SC}}-\epsilon_i\right)^2 \over \Delta^2}
\sim h = {1 \over N} \,,
\end{equation}
where $\phi_k$ is one of the Gaussian wave packets introduced in
Section~\ref{method}, $\psi_i$ and $\epsilon_i$ are the eigenstates and
eigenvalues of the quantum dynamics, and $\epsilon_{i,{\rm SC}}$ are the
semiclassically obtained counterparts to $\epsilon_i$. In other words,
$F_{\phi_k}$ measures the error in the semiclassical eigenvalues, weighing each
eigenvalue error by the overlap of the corresponding eigenstate with $\phi_k$.
A contour plot of $F_{\phi_k}$ versus phase space coordinates $q_k$, $p_k$ is
shown in Fig.~\ref{fig_mixed}~(a), for $N=1/h=256$.

We see that the semiclassical error is peaked in the major stable regions of
phase space, particularly in the large stable island surrounding the $q=p=0$
stable fixed point, and to a somewhat lesser extent in the islands associated
with the period-2 orbit at $p=1/2$. In contrast, $F_{\phi_k}$ remains low in
the region of the chaotic sea, for example, in the vicinity of the unstable
orbit at $q=1/2$, $p=0$. The contour plot in Fig.~\ref{fig_mixed}~(b) shows the
fraction of each wave packet $\phi_k$ consisting of stable trajectories, and
the similarity between the main features in the two parts of the figure
strongly suggests a correspondence between semiclassical accuracy and classical
phase space structure.

The total semiclassical error for a mixed system is of course dominated by the
error associated with the stable regions, and scales in the same way as the
error for a regular system in Section~\ref{regdyn}.

\section{Summary}
\label{summary}

Phase-space semiclassical propagation allows us to make direct comparison
of semiclassical validity in chaotic and stable classical systems. Using
the same semiclassical approximation in both cases results in a semiclassical
error that scales with $\hbar$ in the same way at short times. However,
the growth of the error with time is very different in the two situations.
In the regular case, the error grows coherently because each trajectory
repeatedly visits the same regions of space phase; the mean squared error
therefore grows quadratically with time. In the chaotic case, this coherence
effect does not occur at times short compared with the Heisenberg time,
resulting in a linear growth of the mean squared error.

At the Heisenberg time itself, the mean squared error in the propagator
matrix elements becomes $O(1)$ in the case of a classically stable dynamics,
making it impossible in general to speak of well-defined semiclassical
wave functions or eigenvalues, i.e. ones that are independent of the
choice of semiclassical coordinates. For a given choice of coordinates,
semiclassical quantization generically will produce eigenvalues differing
by the order of a mean level spacing from their quantum counterparts.
Different semiclassical quantizations of a regular system will produce spectra
differing from each other at the same order, making it impossible to uniquely
determine even the order of eigenvalues in the spectrum via semiclassical
methods (unless a particularly favorable set of coordinates can be chosen
where semiclassics happens to be exact, e.g., for a separable dynamics).

In contrast, semiclassical dynamics at the Heisenberg time for a classically
chaotic system becomes increasingly accurate as the system energy is increased.
In the energy domain, the semiclassical error becomes a progressively smaller
fraction of a mean level spacing, so the spectrum can be semiclassically
determined with arbitrarily high accuracy when very highly excited states are
considered. The convergence of semiclassical to quantum behavior for chaotic
system is expected to be independent of the particular semiclassical method
chosen (for example, it is independent of whether a position, momentum,
or phase space semiclassics is used) as long as the methods have the
same scaling with $\hbar$ at fixed time.

All calculations in the present paper were performed for time-dependent
one-dimensional maps, whose scaling properties are equivalent to those of
two-dimensional Hamiltonian systems. In $d=3$ dimensions, or in an interacting
system, the Heisenberg time grows as a higher power of $\hbar^{-1}$ than in the
two-dimensional single-particle case, resulting in a larger accumulated
semiclassical error by the Heisenberg time for both chaotic and regular
systems. For example, the same scaling argument that leads to Eq.~(\ref{eth})
for $d=2$ chaotic systems predicts $O(1)$ semiclassical errors at the
Heisenberg time for chaotic systems, independent of energy, i.e. eigenvalue
errors that remain a constant fraction of a mean level spacing.  In other
words, the breakdown time of the semiclassical approximation will be
proportional to the Heisenberg time in three dimensions, even when the dynamics
is chaotic (and much shorter than the Heisenberg time for regular dynamics).

For $d \ge 4$, e.g., in the case of two interacting particles in two dimensions
with no conserved quantities apart from total energy, the semiclassical
approximation is expected to break down well before the Heisenberg time, even
when the dynamics is fully chaotic. It would be interesting to investigate this
behavior quantitatively for model systems, and also to ascertain how a
higher-order semiclassical approximation may enable semiclassical methods to
remain valid for interacting systems. 

\section*{Acknowledgments} This research was supported in part by the U.S.
Department of Energy, under Grant DE-FG03-00ER41132.

\end{multicols}

\end{document}